\documentclass[aps,twocolumn,amsmath,amssymb, superscriptaddress]{revtex4}
\usepackage[pdftex]{graphicx}
 \usepackage{amsmath}
\usepackage[T1]{fontenc}
\usepackage{amssymb}
\usepackage{amsfonts}
\usepackage{bm}
 \usepackage{amsmath} 
  \usepackage{flushend}
\usepackage{color}
\usepackage{lipsum}
\usepackage{amsfonts} 
\usepackage{amssymb, mathrsfs}
\usepackage{braket}
\usepackage{graphicx} 
\usepackage{subfigure}
\usepackage{bbm}
\def\beq{\begin{equation}}
\def\eeq{\end{equation}}
\def\bsp{\begin{split}}
\def\esp{\end{split}}
\def\bea{\begin{eqnarray}}
\def\eea{\end{eqnarray}}
\def\ba{\begin{array}}
\def\ea{\end{array}}

\def\dg{\dagger}

\def\l.{\left.}
\def\r.{\right.}

\def\ra{\rangle}
\def\la{\langle}

\def\bo{{\vec k}}

\begin{document}

\date{\today}
\title{ Two-dimensional Dirac nodal loop magnons in  collinear antiferromagnets}
\author{S. A. Owerre}
\affiliation{Perimeter Institute for Theoretical Physics, 31 Caroline St. N., Waterloo, Ontario N2L 2Y5, Canada.}


\begin{abstract}
We study the nontrivial linear magnon band crossings in the collinear antiferromagnets on the  two-dimensional (2D) CaVO  lattice, also realized in some iron-based superconductors such as  AFe$_{1.6+x}$Se$_2$ (A = K, Rb, Cs). It is shown that the combination of space-inversion and time-reversal symmetry ($\mathcal{PT}$-symmetry) leads to  doubly-degenerate eight magnon branches, which cross each other linearly along a one-dimensional  loop  in the 2D Brillouin zone.   We show that the Dirac nodal loops (DNLs) are not present in the collinear ferromagnet on this lattice. Thus, the current 2D antiferromagnetic DNLs  are  symmetry-protected and they provide a novel platform to search for their analogs in 2D electronic antiferromagnetic systems.
\end{abstract}
\maketitle

 The study of three-dimensional (3D) topological semimetals in electronic systems such as Dirac \cite{dsm0, dsm1, dsm3, dsm5, dsm2a}, Weyl \cite{wsm0,  wsm3, wsm4}, and nodal-line \cite{ NL1, NL2, NL3,  NL4a, NL5, NL5a} semimetals   has garnered considerable interest in condensed-matter physics. The Dirac and Weyl semimetals feature  linear band crossing points at isolated points in momentum space that are topologically protected. Whereas the nodal-line  semimetals feature symmetry-protected linear band crossing points along a one-dimensional (1D)  loop  in the 3D Brillouin zone (BZ). Indeed, the ubiquitous notion of topological semimetals is not restricted to electronic systems, and can also be applied to the band structures in bosonic systems. One of the areas currently attracting considerable attention  is the topological linear magnon band crossing points in insulating magnetically ordered systems \cite{mw1, mw2,so2}. 

The topological linear magnon band crossing points occur at nonzero energy as a result of the bosonic  nature of magnons. In this respect, realistic topological magnon semimetals with potential practical applications should involve the acoustic (lowest)  magnon branch due to the population effect. By neglecting the Dzyaloshinskii-Moriya (DM)  interaction \cite{dm,dm2} in insulating quantum magnets, DNLs can exist in certain collinear ferromagnets \cite{mok1,so3,per}. Quite distinctively, 3D  collinear \cite{kli}  and noncollinear \cite{so2}  antiferromagnets also allow  3D DNLs even in the presence of the DM interaction. In fact, 3D DNLs are theoretically prevalent in various 3D (layered) compounds when macroscopically symmetry-breaking terms are neglected.  

 Unlike 3D systems, the  2D analogs of DNLs are elusive and much less studied. They do not occur in simple 2D compounds. Recently, a handful of theoretical studies have proposed   2D DNLs in non-magnetic electronic systems with composite lattice structures such as   honeycomb-kagom\'e lattice \cite{lu}, mixed square lattice \cite{yang}, and 2D trilayers \cite{che}. However,  the magnetic analogs of 2D DNLs are still elusive although various 2D magnetically ordered systems exist. Moreover,  2D antiferromagnetic (AFM) systems possess some unique features  that make  them theoretically and experimentally interesting. For instance, AFM systems form the basis of superconductors, and they are currently attracting much attention in the field of spintronics \cite{jung,balt}. Therefore, the realization of  2D  AFM DNLs may play a crucial  role in condensed-matter physics. For  collinear AFM systems, $\mathcal{PT}$-symmetry  guarantees  doubly-degenerate bands, therefore 2D AFM DNLs is not possible in simple ideal 2D honeycomb- and square-lattice  antiferromagnets due to small unit cells.

 In the current work, we circumvent these elusiveness by putting forward a new proposal for 2D AFM DNLs in collinear antiferromagnets  on  the CaVO lattice \cite{rich}, named after the CaV$_4$O$_9$ compound \cite{tan, ueda}, which has a topologically equivalent lattice structure. Such lattice strucure is also realized in  iron-based superconductors such as  AFe$_{1.6+x}$Se$_2$ (A = K, Rb, Cs) \cite{mwang,dai}. The structure of this lattice is equivalent to a decorated (deformed) square lattice with two topologically inequivalent AFM nearest-neighbour bonds. The collinear antiferromagnet state on this lattice features eight sites in the unit cell. We show that the eight magnon branches are doubly-degenerate due to $\mathcal{PT}$-symmetry. Interestingly, they feature linear magnon band crossings  along a 1D loop  in the 2D  BZ. We show that the loops of Dirac nodes  are not present in the collinear ferromagnet on this lattice. Therefore, the 2D AFM DNLs are symmetry-protected and provide the first concrete example in 2D AFM systems.

\begin{figure}
\centering
\includegraphics[width=1\linewidth]{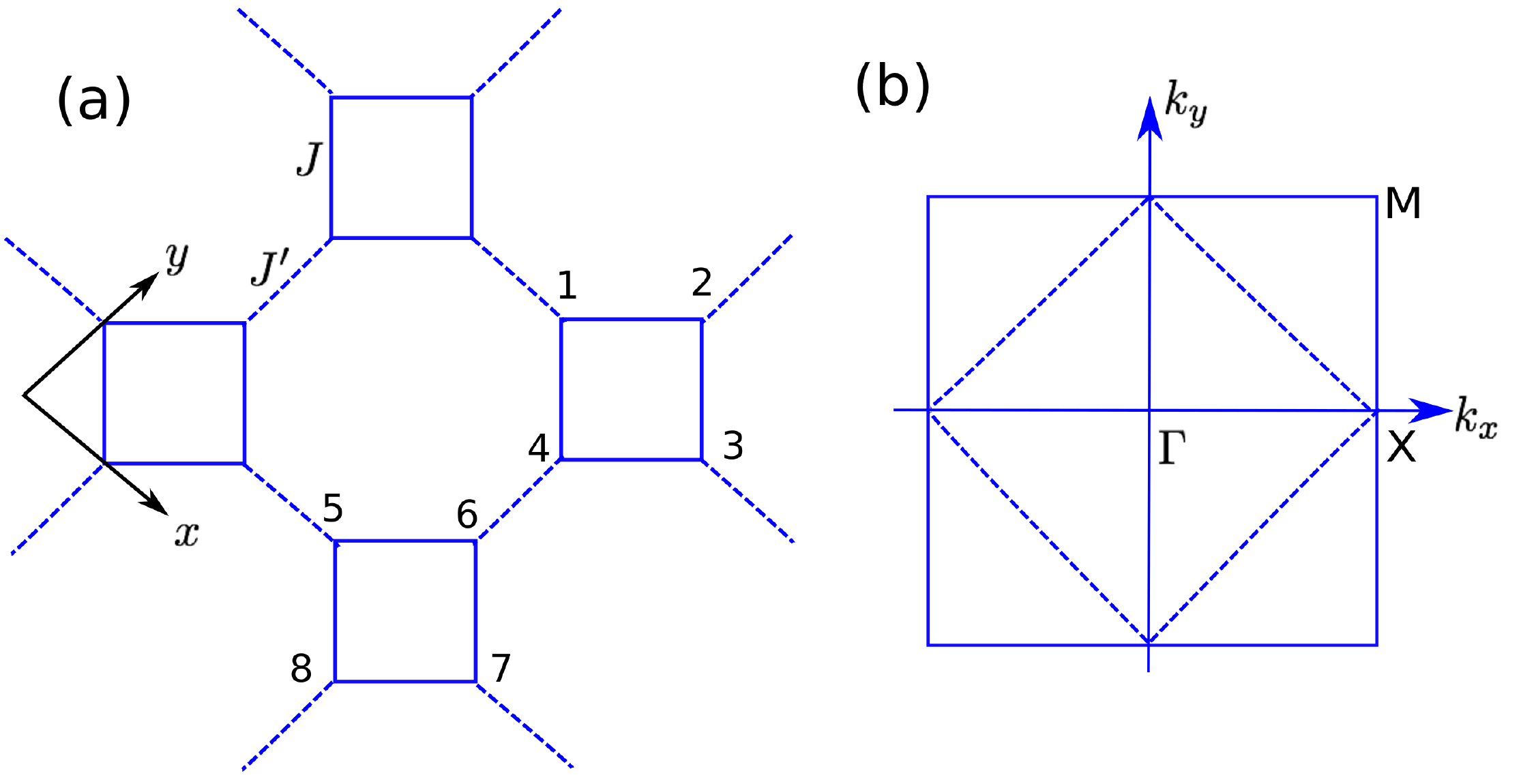}
\caption{(a) The CaVO lattice with two topologically inequivalent AFM bonds indicated by $J$ and $J^\prime$. The numbers label the 8-site AFM unit cell. (b) The corresponding square lattice  BZ with high symmetry points indicated. The dashed square represents the AFM BZ.  }
\label{CaVO}
\end{figure}

We study the  Heisenberg antiferromagnets  on  the CaVO lattice
\begin{align}
  \mathcal H=J\sum_{\la ij\ra}\vec S_{i}\cdot {\vec S}_{j}+J^\prime\sum_{\la ij\ra^{\prime}}\vec S_{i}\cdot {\vec S}_{j}.
\end{align}
There are two topologically inequivalent AFM bonds, namely $J$ and $J^\prime$. The first summation $\la ij\ra$ is taken over nearest-neighbour sites on the intralayer  four-spin plaquettes, and the second summation $\la ij\ra^{\prime}$ is taken over nearest-neighbour sites on the interlayer  dimer bonds as depicted in  Fig.~\ref{CaVO}a.  We note that  there is no geometrical spin frustration on this lattice for a wide range of $J^\prime/J$ \cite{moo}. Thus, the spins form a perfect collinear AFM structure with $\mathcal{PT}$ symmetry.  The perfect collinear AFM structure is accompanied by the presence of an inversion center, therefore the existence of the DM interaction is forbidden on the CaVO lattice. There are four sites per structural unit cell, but  the AFM unit cell contains eight sites, {\it i.e.} doubled as shown in Fig.~\ref{CaVO}(a). The CaVO lattice is also bipartite, and thus can be divided into two sublattices A and B. 

We now study the magnetic excitations above the classical N\'eel ground state. For this purpose, it is expedient to introduce the Holstein Primakoff  bosons:  
\begin{align}
&S_{i}^{ z}= S-a_{i}^\dagger a_{i},~S_{i}^+= \sqrt{2S}a_{i}=(S_{i}^-)^\dg,
\end{align}
for up pointing spins, 
\begin{align}
  S_{j}^{ z}= -S+b_{j}^\dagger b_{j},~S_{j}^+= \sqrt{2S}b_{j}^\dg=(S_{j}^-)^\dg,
  \end{align}
  for down pointing spins. Here $a_{i}^\dagger (b_{i})$ are the bosonic creation (annihilation) operators on sublattice A (B), and  $S^\pm_{j}= S^x_{j} \pm i S^y_{j}$ denote the spin raising and lowering  operators. 
The bosonic hopping model is given by  

  \begin{align}
  \mathcal H_{\rm{sw}}&= S\sum_{\la ij\ra}J_{ij}\big[(a_{i}a_{i}^\dg + b_{j}b_{j}^\dg)+(a_{i}^\dg b_{j}^\dg+\rm{h.c.})\big],
  \end{align}
where $J_{ij}=J,J^\prime$. The   Fourier transformed Hamiltonian  into momentum space yields 
 $\mathcal H_{\rm{sw}}=\sum_\bo \Psi^\dg_\bo \mathcal{H}(\bo)  \Psi_\bo,$ where
\begin{align}
\mathcal{H}(\bo)= 
\begin{pmatrix}
\mathcal{H}_{+}(\bo) & 0\\
0&\mathcal{H}_{-}(\bo)
\end{pmatrix}.
\label{ham}
\end{align}
 The  basis vector is given by $\Psi^\dg_\bo=(\psi^\dg_\bo, \psi_{-\bo})$, with $\psi^\dg_\bo=(a_{\bo 1}^{\dg},\thinspace a_{\bo 2}^{\dg},\thinspace a_{\bo 3}^{\dg}, \thinspace a_{\bo 4}^{\dg},b_{-\bo 5},\thinspace b_{-\bo 6},\thinspace b_{-\bo 7}, \thinspace b_{-\bo 8})$. We note that each block Hamiltonian is an $8\times 8$ Hermitian matrix representing $S_z=+1$ and $S_z=-1$ spin sectors that satisfies  $\mathcal{H}_{-}(\bo)=\mathcal{H}_{+}^*(-\bo)$.  Therefore, $\mathcal{H}(\bo)$ is invariant under $\mathcal{PT}$ symmetry given by $\mathcal{PT}=\sigma_x\otimes\sigma_0 \mathcal{K}$, where $\mathcal{K}$ is complex conjugation, and $\sigma_x$ is a Pauli matrix with an identity $\sigma_0$. The Hamiltonian for the $S_z=+1$ sector is given by
\begin{align}
\mathcal{H}_{+}(\bo)= 
\begin{pmatrix}
\mathcal{A}(\bo) & \mathcal{B}(\bo)\\
\mathcal{B}^*(-\bo)&\mathcal{A}(\bo)^\dg
\end{pmatrix},
\label{ham}
\end{align}
with $\mathcal{A}(\bo)=S(2J+J^\prime)\rm{I}_{4\times 4}$, and $\mathcal{B}(\bo)$ is succinctly given by
\begin{align}
\mathcal{B}(\bo)=S& 
\begin{pmatrix}
0&J&J^\prime  e^{ik_y}&J\\
J&0&J&J^\prime e^{ik_x}\\
J^\prime   e^{-ik_y}&J&0&J\\
J&J^\prime  e^{-ik_x}&J&0
\end{pmatrix},
\end{align}
where $\rm{I}_{4\times 4}$ is the identity $4\times 4$ matrix. The spin wave Hamiltonian $\mathcal{H}(\bo)$ can be diagonalized by the Bogoliubov transformation, and there are eight doubly-degenerate magnon branches due to  $\mathcal{PT}$ symmetry. 

 \begin{figure}
\centering
\includegraphics[width=1\linewidth]{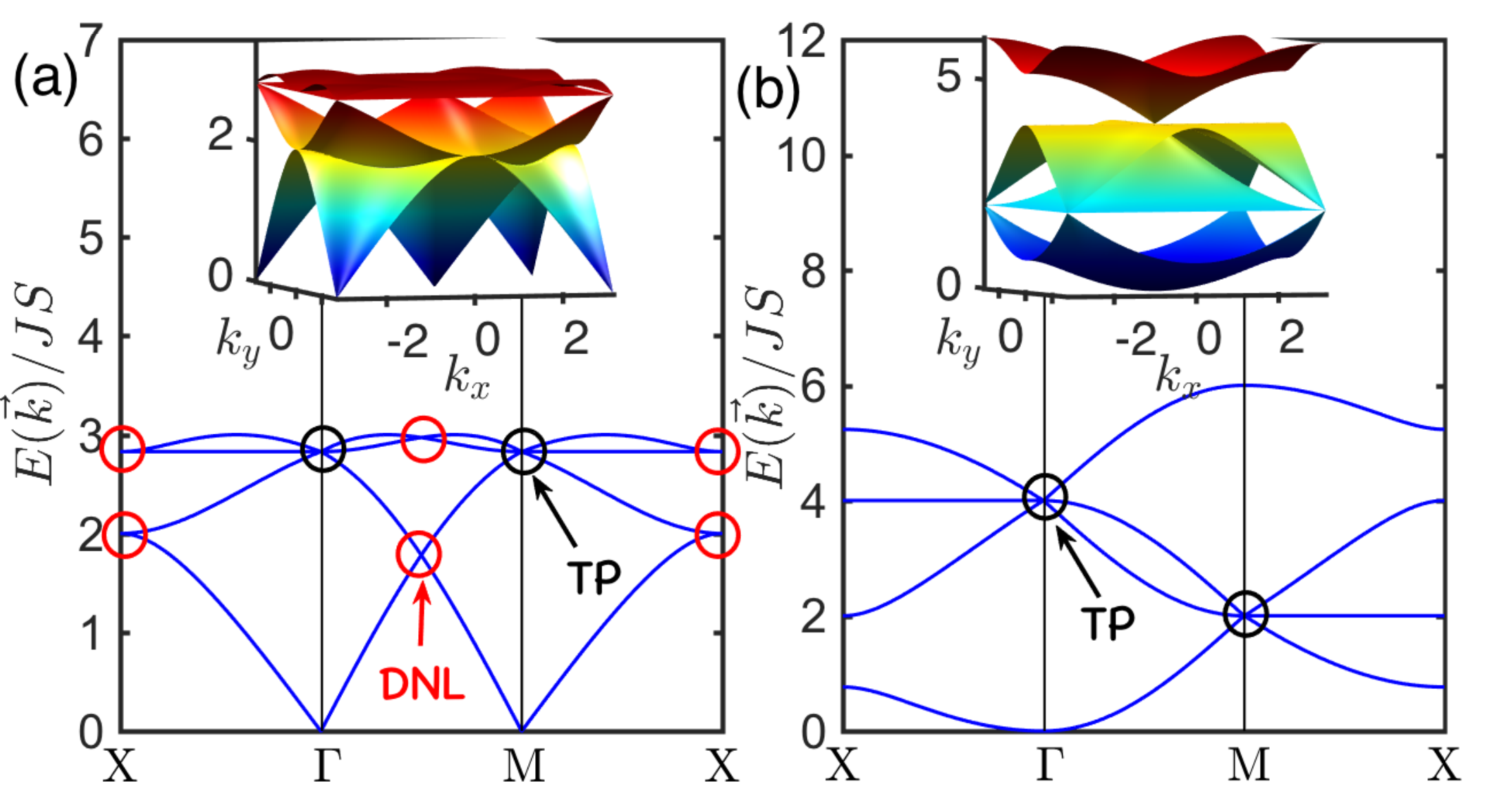}
\includegraphics[width=1\linewidth]{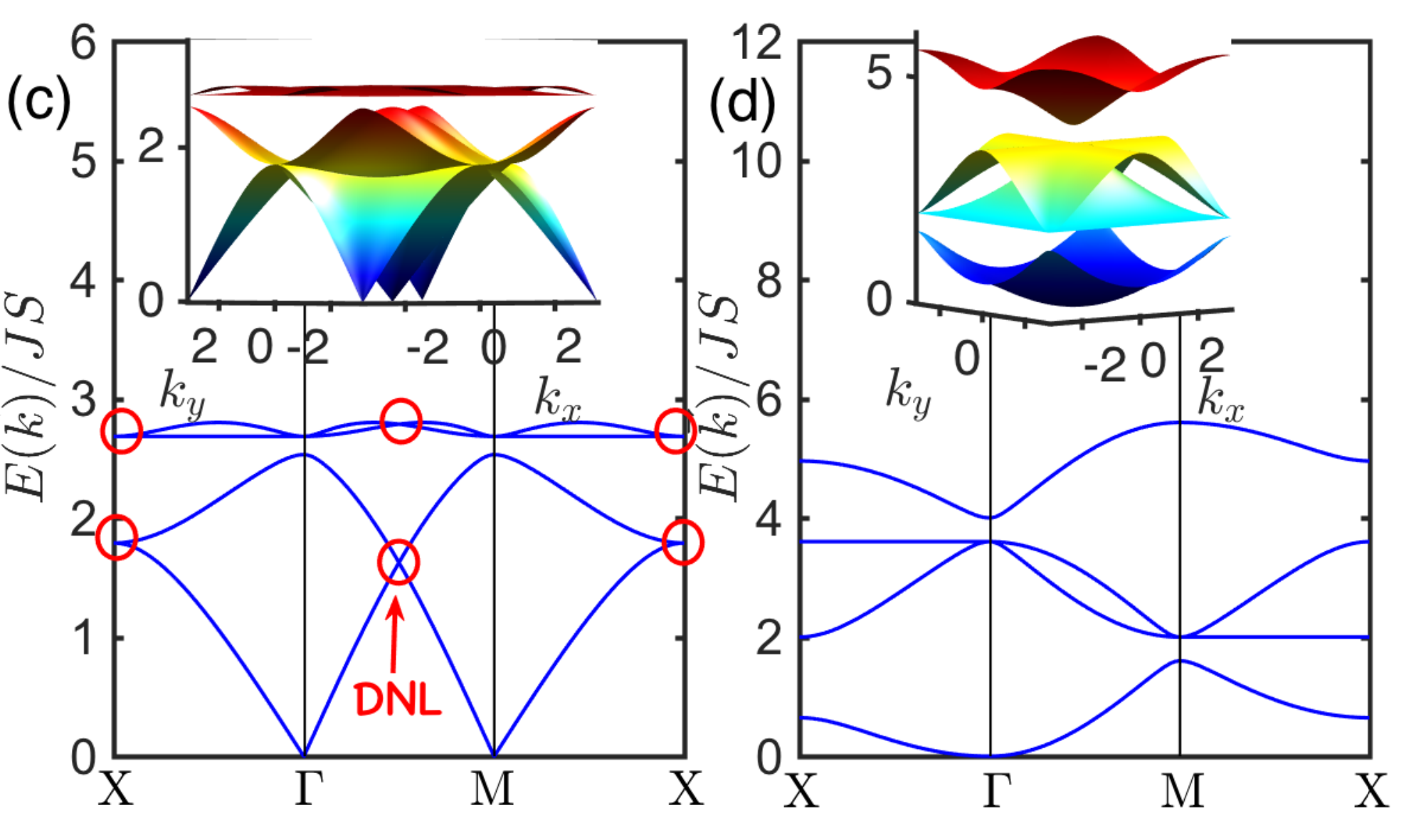}
\caption{(a,c) The  doubly-degenerate eight magnon branches in the collinear antiferromagnet on the CaVO lattice for $J^\prime/J=1$ and $J^\prime/J=0.8$ respectively.  (b,d)  The  nondegenerate four magnon branches in the collinear ferromagnet on the CaVO lattice for $J^\prime/J=1$ and $J^\prime/J=0.8$ respectively. The four-fold degenerate DNLs are indicated by the red circles, and the magnon triple points (TPs) (six-fold degenerate in the AFM phase and three-fold degenerate in the FM phase) are indicated by the black circles. Insets show the respective 3D magnon bands. }
\label{DNL}
\end{figure}

 \begin{figure}
\centering
\includegraphics[width=1\linewidth]{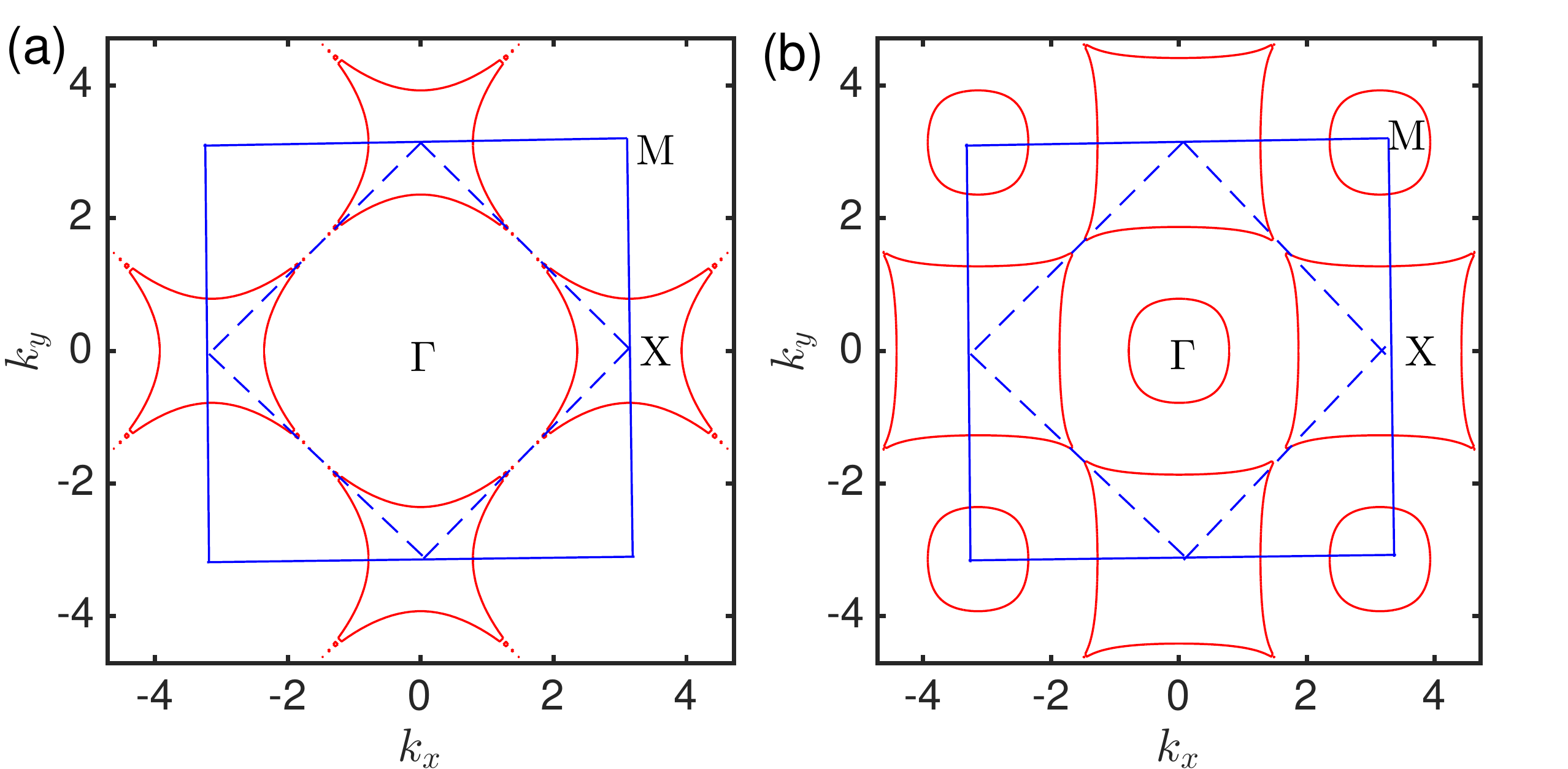}
\caption{Loops of Dirac magnon nodes formed by (a) the acoustic magnon band crossing and  (b) the optical magnon band crossing.  The solid rectangles denote the structural  BZ with high symmetry points indicated, and the dashed squares represent the AFM BZ zone. We set $J^\prime/J=1$ corresponding to Fig.~\ref{DNL}(a). }
\label{loop}
\end{figure}

 We now study the magnon band structure of collinear antiferromagnet on the CaVO lattice. For the purpose of comparison, we will also show the magnon band structure of corresponding collinear ferromagnet. These are shown in Figs.~\ref{DNL}(a) and \ref{DNL}(b) respectively for $J^\prime/J=1$. In the collinear AFM case in Fig.~\ref{DNL}(a), there are eight doubly-degenerate magnon branches. The magnon bands show linear Goldstone modes at the $\Gamma$ and the M points of the BZ as expected in systems with a square lattice structure. Remarkably, the doubly-degenerate acoustic magnon band and the doubly-degenerate optical magnon bands  feature a line of Dirac magnon nodes along the  $\Gamma$-M line. The X point also contributes to  the line of Dirac magnon nodes, with energy very close  to those along the  $\Gamma$-M line. Hence, the four-fold degenerate magnon band crossing forms a 1D curve as depicted in the inset of Fig.~\ref{DNL}(a).   Moreover, there also exists  2D magnon triple points (TPs) (six-fold degenerate) at the $\Gamma$ and the M points. The TPs are isolated points and do not form line of Dirac magnon nodes.

For comparison, we have also shown the magnon band structures of the corresponding collinear ferromagnet in Fig.~\ref{DNL}(b). In this case, there are only four sites in the unit cell and the four magnon branches are nondegenerate. In contrast to collinear antiferromagnets, there are no line of Dirac magnon nodes in the ferromagnetic system, however the TPs (three-fold degenerate in this case) are still present at the $\Gamma$ and the M points, but at different energies.  The existence of 2D TPs in both ferromagnets and antiferromagnets is due to $\mathcal{C}_4$ symmetry of the square lattice,  and they are immediately gapped out  for $J\neq J^\prime$ \footnote{Unlike 2D systems, 3D magnon TPs are robust against magnetic anisotropy as well as the DM interaction, see Ref. \cite{soltp}} as shown in Figs.~\ref{DNL}(c) and \ref{DNL}(d). In contrast, the 2D AFM line of Dirac magnon nodes are not a consequence of $\mathcal{C}_4$ symmetry and remain intact for $J\neq J^\prime$, hence they are symmetry-protected by $\mathcal{PT}$ symmetry of the collinear N\'eel structure. 

Furthermore, the 2D linear magnon band crossings also form 1D loops in the 2D BZ plane centred at $\Gamma$, M and X points as shown in  Figs.~\ref{loop}(a) and \ref{loop}(b).  It is noted that the TPs have no nodal-loop magnons in the 2D BZ as they are point nodes.  Although the 2D AFM DNLs are symmetry-protected, we can further confirm their topological protection from the parity eigenvalues at the four time-reversal-invariant momenta in two dimensions given by $\Gamma_i=\lbrace(0,0),~(0,\pi),~(\pi,0),~(\pi,\pi)\rbrace$.  The $\mathbb Z_2$ invariance $\nu$ is given by \cite{fu} $(-1)^{\nu}=\prod_{i=1}^4\prod_{n=1}^{N}\xi_{n}(\Gamma_i)$, where  $\xi_{n}(\Gamma_i)$ is the parity eigenvalue associated with the magnon bands that form DNLs. Numerical estimate shows that $\nu=1$ which yields a nonzero topological invariant $\mathbb Z_2=1$, confirming the  topological protection of the DNLs.

 In a nutshell, we have shown that symmetry-protected  2D AFM DNLs exist in insulating collinear antiferromagnets on the CaVO lattice. The current 2D AFM DNLs occur at the acoustic magnon branch, which is very crucial in realistic magnetic materials due to the population effect of bosons at low temperatures. Remarkably, the bosonic excitations could provide the realization of 2D AFM DNLs.  They can be probed experimentally using the inelastic neutron scattering by synthesizing the right 2D AFM materials. The current results also provide a novel platform to search for 2D AFM DNLs in electronic systems.   As we mentioned above,   2D AFM DNLs do not occur in the simple ideal 2D quantum antiferromagnets. For instance, the insulating collinear antiferromagnets on the square and honeycomb lattices have a single doubly-degenerate magnon branch, which cannot possess linear band crossing between two magnon branches  in the BZ. The proper inclusion of the DM interaction  should lead to spin canting and noncollinear spin structure. However, a small DM interaction will not completely eliminate the DNLs in the entire BZ due to the high symmetry protection of the lattice. Since there is no experimental evidence of the DM interaction on the current lattice, we believe that the DNLs should be present and robust. Upon the completion of this paper, we became aware of a recent arXiv preprint  \cite{wang}, where 2D DNLs have been proposed in an electronic AFM system, but with nonsymmorphic analogue symmetry as opposed to $\mathcal{PT}$ symmetry.  

 Research at Perimeter Institute is supported by the Government of Canada through Industry Canada and by the Province of Ontario through the Ministry of Research and Innovation.

\end{document}